\documentclass[aps,prl,fleqn,twocolumn,superscriptaddress,floatfix]{revtex4}

\usepackage{epsfig}
\usepackage{amsmath,amssymb,amsfonts}

\begin{document}


\title{Centrality Dependence of Direct Photons in Au+Au Collisions at 
$\mathbf{\sqrt{s_{NN}}=200}$ GeV}

\author{Rainer~J.~Fries}
\affiliation{School of Physics and Astronomy, University of Minnesota,
             Minneapolis, MN 55455}

\author{Berndt~M\"uller}
\affiliation{Department of Physics, Duke University, Durham, NC 27708}

\author{Dinesh~K.~Srivastava}
\affiliation{Variable Energy Cyclotron Center, 1/AF Bidhan Nagar, 
             Kolkata 700 064, India}

\preprint{NUC-MINN-05/6-T}

\date{\today}

\begin{abstract}
We calculate the spectra of high energy photons emitted in relativistic
Au+Au collisions for various centralities and compare to data recently 
 collected at the Relativistic Heavy Ion Collider by the PHENIX collaboration. 
Our results for photons from primary hard scatterings and photons from 
interactions of jets with the medium are consistent with the measurements 
of neutral pion and direct photon production in $p+p$ collisions and
give a good description of direct photon 
spectra measured in Au+Au collisions. The contribution of photons from 
jet-to-photon conversion in the medium can be as large as the photon yield
from hard scatterings in the momentum range $p_T \approx 2\ldots 6$ GeV/$c$. We
show that this novel mechanism is not ruled out by any existing data.
\end{abstract}

\maketitle


High energy nuclear collisions are studied to understand the properties
of strongly interacting matter at the highest energy densities and to
search for the transition of hadronic matter to a quark-gluon 
plasma \cite{Harris:1996zx}. Electromagnetic probes are excellent tools to 
gather information about the dense and hot medium created in such collisions.
The mean free path of photons is much larger than the typical size of 
the emerging fireball, allowing them to escape nearly unperturbed once 
created in the system \cite{Feinb:76}. Since photons are emitted throughout 
the history of the fireball, one expects to obtain information that is 
complementary to that gained from the measurement of hadrons, which are 
subject to strong final-state interactions and are reflecting 
the latest stage of the collision.

Thermal radiation of photons can occur both from a quark-gluon plasma 
\cite{KaLiSei:91,AMY:01}
and from a hot hadronic 
gas \cite{XSB:92,Turb:04}.
Because the photon spectrum 
reflects the thermal distribution of the emitting matter, it can serve as 
a thermometer. On the other hand, high energy photons are 
most likely to be created either in hard processes between 
partons in the colliding nuclei \cite{Owens:86}, or in final-state 
interactions of high energy jets \cite{FMS:02}. Photons from jets
naturally carry information about the medium they traverse and hence
can serve as a medium probe.

The latter process was introduced in our earlier work \cite{FMS:02} where 
we predicted the existence of a medium induced component in the photon 
spectrum in the intermediate $p_T$ range. In this novel mechanism, a hard 
quark passing through the medium converts into a photon carrying most of
its transverse momentum. Recently Zakharov has developed the related concept 
of modified photon bremsstrahlung of jets in the medium \cite{Zakharov:04}. 
An up-to-date calculation of photon emission including such jet-medium 
interactions can be found in the work of Turbide {\it et al.} \cite{TGJM:05}.
The general arguments given for photons in this work also apply to the 
emission of lepton pairs initiated by quark jets in the plasma \cite{SGF:02}.

Relativistic nuclear collisions at different centralities result in the
formation of systems of different sizes, initial temperatures and life 
times. Assuming that the yield of processes involving large momentum transfer 
(hard processes) scale as the number of collisions $N_\mathrm{coll}$, while 
those with small momentum transfer (soft processes) scale as the number of
participants $N_\mathrm{part}$, one 
can predict the relative importance of the two contributions as a function
of centrality (see e.g.~\cite{many}). The recently published data on 
single photon production in $p+p$ collisions \cite{phenix:05gammapp} at 
$\sqrt{s_{\rm{NN}}}=200$ GeV \cite{phenix:05gammapp} and the centrality dependence 
of the photon yield in Au+Au collisions measured by PHENIX at 
the Relativistic Heavy Ion Collider (RHIC) \cite{phenix:05gammaauau} is our first 
chance to obtain an understanding of the different photon sources.

Here, we shall address the question whether the data allow us 
to draw conclusions about the relative contributions from different photon
production mechanisms to the direct photon spectrum in Au+Au 
collisions. The data is still limited by (statistical and systematic)
uncertainties, but the results 
represent a remarkable achievement because of the
experimental challenge to subtract the background of
photons from secondary hadronic decays, especially neutral pions.
Fortunately, in Au+Au collisions at RHIC this background is 
reduced at intermediate and high transverse momenta $p_T > 2$ GeV/$c$ 
by the large final state suppression (commonly called ``jet quenching'') 
for pions and other hadrons \cite{phenix:03jq,star:03jq}.
Thus, while the low-$p_T$ part of the photon spectrum remains clouded by 
the large 
hadronic background, this is an invitation to look for electromagnetic
signals of a quark gluon plasma at intermediate and high 
$p_T$ \cite{FMS:02,GAFS:04}.


Let us briefly recall the photon sources that are important at $p_T > 2$ 
GeV/$c$. We define prompt photons as those coming from hard Compton and 
annihilation processes (e.g.\ $q+g\to q+\gamma$ and $q+\bar q\to g+\gamma$) 
and those emitted from jets after the hard interaction as bremsstrahlung 
(e.g.\ $q\to q+\gamma$). A baseline for these processes can be obtained 
by studying $p+p$ and $p+\bar p$ interactions. Besides an obvious scaling 
from $p+p$ collisions to Au+Au collisions with the number $N_{\rm coll}$,
and an isospin correction due to the presence of $p+n$ and $n+n$ collisions, 
Compton and 
annihilation spectra are expected to be altered only marginally by
initial state nuclear effects.
Neutral pion spectra measured in $d$+Au collisions indicate that such 
effects from shadowing and $k_T$-broadening are small
\cite{phenix:03pidau}. The hard process itself is pointlike on typical 
scales of the medium and therefore remains unaffected. Direct photon spectra
measured in $d$+Au could help to quantify the role of initial state effects.

Bremsstrahlung emission from jets, on the other hand, takes place on more 
extended distances and thus can be affected by final state interactions. 
Energy loss effects would then be noticeable as a suppression of bremsstrahlung
\cite{JJMS:02}, while secondary scattering in the medium could initiate 
additional bremsstrahlung or modify bremsstrahlung by coherence effects 
\cite{Zakharov:04}. In our work we assume that the total photon yield from 
jets can be described by the superposition of the usual vacuum bremsstrahlung 
with an additional contribution from jets annihilating and Compton scattering 
in the medium as described in \cite{FMS:02}.

Photons from secondary hard scatterings in the pre-equilibrium phase have 
been studied elsewhere and could contribute to photons at intermediate and 
high $p_T$ \cite{BDRS:97}. In order to minimize possible effects
from neglecting pre-equilibrium emission, we assume a very early 
thermalization time, $\tau_0 \approx 0.15$ fm/$c$, in central collisions, 
not much larger than the formation time of minijets. An early thermalization 
of the matter created in Au+Au collisions at RHIC is supported by several 
observations, especially the large elliptic flow. We also assume that the 
matter is chemically equilibrated at the initial time. Secondary collisions
are then either between thermalized partons or between hard scattered 
partons and the medium, which are both taken into account here. Similar 
strategies for dealing with the uncertainties of pre-equilibrium 
emission have been invoked elsewhere \cite{dEPe:05}.


We now recall the specifics of the jet-medium interaction process. We 
denote the momenta of the jet, the thermal parton, and the photon by 
${\mathbf p}_{\rm jet}$, ${\mathbf p}_{\rm th}$ and ${\mathbf p}_{\gamma}$, 
respectively. The leading order QCD Compton and annihilation cross sections 
are peaked in the forward and backward directions. In the laboratory frame 
we have $|{\mathbf p}_{\rm jet}| \gg |{\mathbf p}_{\rm th}| \sim T$, where 
$T$ is the temperature of the plasma. For high energy photons, 
i.e.\ $|{\mathbf p}_{\gamma}| \gg T$, this implies that 
$|{\mathbf p}_{\gamma}| \approx |{\mathbf p}_{\rm jet}|$. This justifies
calling the process a conversion of a jet (or more precisely: a quark
initiating a jet) into a photon with similar momentum \cite{FMS:02}. 
Only quark jets contribute to this process, because the gluon-photon 
scattering cross section is not peaked at forward angles.

The rate of photon production by annihilation and Compton scattering of 
jets in the medium can be approximated as \cite{FMS:02}
\begin{equation}
\label{eq:1}
  \frac{E_\gamma d N}{d^3 {\mathbf p}_\gamma d^4 x} =
  \frac{\alpha\alpha_s}{4\pi^2} \sum_q e_q^2 f_q ({\mathbf p}_\gamma,x)
  T^2 \left[ \ln \frac{4 E_\gamma T}{m^2_{\rm th}} + C \right] 
\end{equation}
where $C=-1.916$, $m^2_{\rm th} = g^2 T^2/6$ and $\alpha_s = g^2/(4\pi)$ is 
the strong and $\alpha$ the electromagnetic coupling. $q$ denotes all light
quark and antiquark species with charge $e_q$, and $f_q$ is the distribution
of minijets of flavor $q$. It is worth emphasizing that the conversion 
property of the process is reflected in (\ref{eq:1}) by the fact that 
the photon spectrum is directly proportional to the jet spectrum $f_q$.
Calculations without some of the the simplifying approximations~\cite{FMS:02}
that lead to the expression in Eq.~(\ref{eq:1}) produce only small
modifications of the photon yield \cite{TGJM:05}.

In order to calculate the yield from thermal emission and jet
conversion, we need to model the space-time evolution of the fireball. 
We want to compare with data taken at midrapidity, hence we 
assume the boost invariant Bjorken scenario for the longitudinal expansion.
In addition we neglect the marginal transverse expansion of the
plasma on the jet-photon conversion part, i.e.\ during the short time 
jets are propagating through the QGP.
But the transverse expansion is fully taken into account for
the thermal contribution. 

We integrate over the conversion vertex $x$ in (\ref{eq:1}) in terms of the 
space-time rapidity $\eta$, the radial coordinate $\mathbf r$ 
(with $r=|{\mathbf r}|$) at which the jet is emitted, and the proper time
$\tau$, so that $d^4 x = \tau d\tau d\eta d^2 r$.
We parameterize momenta ${\mathbf p}$ by their
rapidity $y$ and transverse momentum ${\mathbf p}_T$, 
($p_T = |{\mathbf p}_T|$).
Starting from the yield $dN_{\rm jet}/(d^2 p_T dy)$ of quarks initiating a
jet, the distribution of hard partons in a boost-invariant 
scenario is given by \cite{LinGyu:94,FMS:02}
\begin{equation}
  \label{eq:2}
  f_q({\mathbf p},x) = \frac{(2\pi)^3}{g_q \pi R_T^2 \tau p_T} 
  \frac{d N_{\rm jet}}{d^2 p_T dy} \rho({\mathbf r}) \delta(y-\eta)
\end{equation}
where $R_T$ is the radius of the fireball, $g_q=6$ the spin-color degeneracy 
factor of parton $q$, and $\rho$ is a transverse profile function normalized
to one. For simplicity, we will take the profile to be azimuthally symmetric 
with area density $\rho(r) = 2(1-r^2/R_T^2) \Theta(R_T-r)$ and transverse 
radius $R_T=1.2(N_{\rm part}/2)^{1/3}$.

The number of participants, 
$N_{\rm part}$, is a function of centrality, but we neglect the azimuthal 
asymmetry in noncentral collisions. This approximation does not allow us to 
predict the azimuthal distribution of photons, but it 
introduces only a small error for the yield integrated over the azimuthal 
angle. 
The jet spectrum for an arbitrary impact parameter $b$ is obtained by scaling
\begin{equation}
\frac{d N_{\rm jet}}{d^2 p_T dy} (b)=\frac{T_{\rm AA}(b)}{T_{\rm AA}(0)}
\times \frac{d N_{\rm jet}}{d^2 p_T dy} (b=0).
\end{equation}
This expression shows that the photon yield due to jet-photon conversion 
scales as a product of the nuclear thickness and a term dependent on the 
initial conditions, like temperature and size, and on the history of the 
evolution of the fireball. A measurement of the photon spectra at different 
centralities could thus help reveal this information, as the primary hard 
photons 
will scale with the nuclear thickness $T_{AA}$ alone.


We use a leading order perturbative calculation of jets and of direct
Compton and annihilation photons \cite{Owens:86}, using CTEQ5L parton 
distributions \cite{cteq5:99}. For gold nuclei EKS98 corrections of the
nuclear parton distributions \cite{EKS:98} are taken into account. 
Higher order perturbative effects are accounted for by a phenomenological 
$K$-factor, which is known to depend weakly on the transverse momentum $p_T$ 
at RHIC energies \cite{BFLPZ:00} and is here assumed to be constant.

\begin{figure}
  \begin{center}
  \epsfig{file=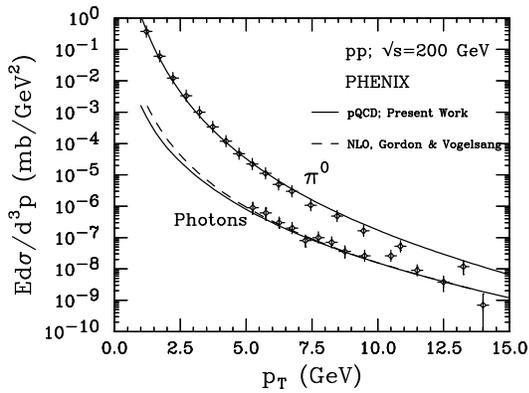,width=0.8\columnwidth}
  \end{center}
  \caption{\label{fig:1} Invariant cross section $Ed\sigma/d^3p$ for $\pi^0$ 
   and $\gamma$ production in $p+p$ collisions at $\sqrt{s} = 200$ GeV as 
   a function of transverse momentum $p_T$ at $y=0$. Data points are from
   the PHENIX experiment \cite{phenix:03pi0pp,phenix:05gammapp}. We also
   show a full NLO calculation \cite{GoVo:93}.}
\end{figure}

Vacuum photon bremsstrahlung is obtained by convoluting the jet distributions
with the photon fragmentation functions given in \cite{Owens:86} ($\Lambda
= 200$ MeV), and we use KKP fragmentation functions \cite{BKK:00} for the
yield of neutral pions. We calculate the $\pi^0$ and photon cross sections 
at midrapidity for $p+p$ and compare to the values measured by PHENIX. This 
allows us to fix the $K$ factors for jet production, $K_{\rm jet}$, and direct
Compton and annihilation photons, $K_{\rm \gamma}$, independently. 
We obtain a good description of the $p+p$ data with $K_{\rm jet}=1.8$
and $K_{\rm \gamma}=1.5$, see Fig.~\ref{fig:1}. These values agree with 
those reported in ref.~\cite{TGJM:05} on the basis of a next-to-leading
order (NLO) perturbative QCD calculation and our results also match the NLO 
calculation by Gordon and Vogelsang \cite{GoVo:93}.

Once the $K$-factors are fixed we apply them to calculate the yield of 
direct photons in Au+Au collisions. 
The prompt photon yield in Au+Au is close to that in $p+p$ scaled with 
$T_{AA}$, due to small shadowing corrections. Note, however, that there is an 
effect from the admixture of $p+n$ and $n+n$ collisions with
different photon yields. 
Therefore a simple scaling of $p+p$ results, as often invoked by the 
experimental collaborations for comparison, introduces a small but systematic 
bias of the photon yield from hard processes towards larger values.

\begin{table}
\begin{center}
\begin{tabular}{c||r|r|r|r|r}
  Centrality & 0-10\% & 10-20\% & 20-30\% & 30-40\% & 40-50\% \\
  \hline\hline
  $N_{\rm part}$ & 326.2 & 234.5 & 166.0 & 114.0 & 75.0 \\
  $T_{AA}$ [1/mb] & 22.75 & 14.35 & 8.00 & 5.23 & 2.86 \\
  $\tau_0$ [fm/$c$] & 0.151 & 0.166 & 0.182 & 0.201 & 0.221 \\
  $T_0$ [MeV] & 0.434 & 0.395 & 0.361 & 0.328 & 0.297 \\
  \hline
\end{tabular}
\caption{\label{tab:1} Number of participants $N_{\rm part}$, nuclear
thickness factor $T_{AA}$, initial time $\tau_0$ and initial temperature
$T_0$ for different centrality bins.}
\end{center}
\end{table}

Next we use the jet yields in Au+Au to calculate the photons from jet-photon
conversions using (\ref{eq:1}) and (\ref{eq:2}). For the fireball we assume 
a thermally and chemically equilibrated plasma at some early time $\tau_0$ 
with initial temperature $T_0$. The product $T^3 \tau$ is conserved during 
the isentropic longitudinal expansion, and its value can be fixed by the total 
hadron multiplicity $dN_h/dy$. The initial conditions are determined by 
imposing the thermalization condition $\tau_0 \approx 1/(3T_0)$. Table 
\ref{tab:1} lists the initial times and temperatures inferred from the 
charged hadron multiplicity $dN/d\eta$ measured by PHENIX \cite{phenix:03jq} 
for different centrality bins. Nuclear thickness factors and number of 
participants are also given.

The transverse profile of the initial temperature is fixed by assuming that
the energy density scales with the square of the nuclear thickness
factors, so that $T_i(r) = T_0\left[2(1-r^2/R_T^2)\right]^{1/4}$.
This has been found to work well for the study of the centrality 
dependence of hadronic spectra
~\cite{dks1} as well as lepton pair production~\cite{dks2}, and should be
sufficient for our purpose.
Starting from the initial time $\tau_0$, we integrate (\ref{eq:1}) to 
either the time $\tau_f$, when the medium has cooled to the critical 
temperature $T_f = 160$ MeV, or to the time when the jet reaches the 
boundary of the fireball, whichever comes first.

For the calculation of thermal photons we assume that the thermally and 
chemically equilibrated quark-gluon plasma, which undergoes 
a boost-invariant longitudinal and an azimuthally symmetric transverse 
expansion~\cite{dks3}, converts to a hadronic gas below $T_c$. 
The production of photons from the deconfined phase is calculated using 
the complete leading order emission rate \cite{AMY:01}. The latest results 
of Turbide {\em et al.} \cite{Turb:04} for photon emission from hadronic 
matter are used.

\begin{figure}
  \begin{center}
  \epsfig{file=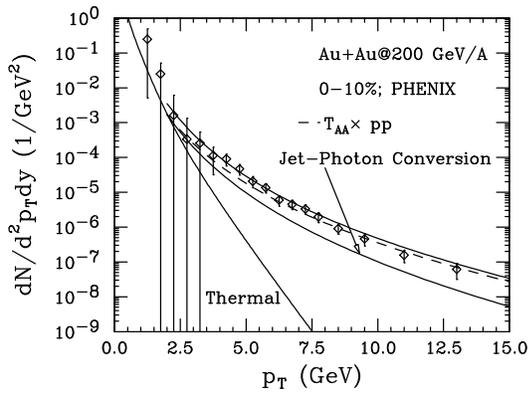,width=0.8\columnwidth}
  \end{center}
  \caption{\label{fig:2} Photon yield $dN/(d^2p_T dy)$ as a function of $p_T$
    for $y=0$ in central (0-10\%) Au+Au collisions at $\sqrt{s}=200$
    GeV. We show primary hard photons (dashed), jet-photon conversion 
    (solid and labelled) 
    and the sum of both (upper-most solid curve). Thermal photons are also 
    shown (solid and labelled). Data are from the 
    PHENIX collaboration \cite{phenix:05gammaauau}.}
\end{figure}

\begin{figure}
  \begin{center}
  \epsfig{file=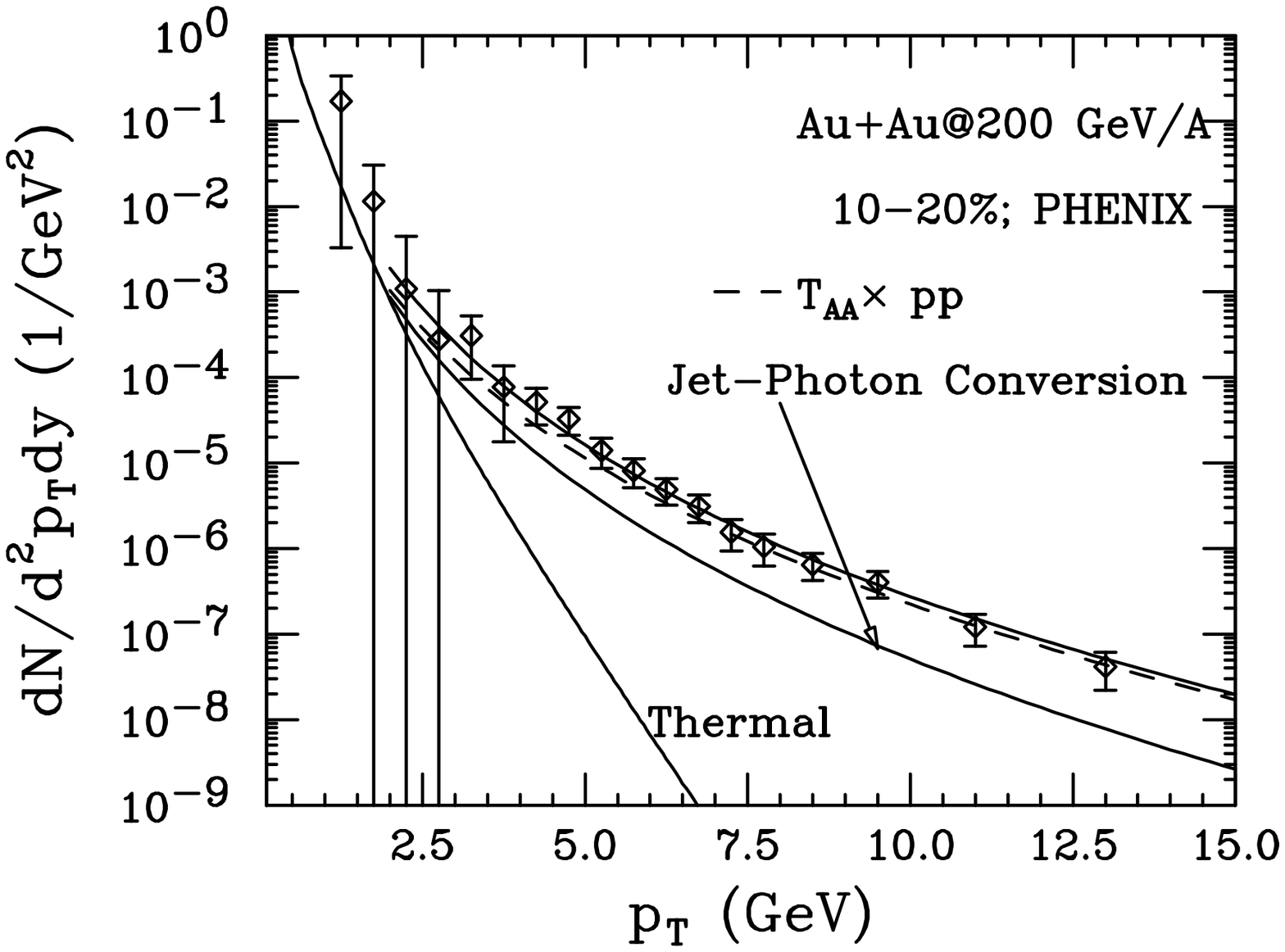,width=0.8\columnwidth}
  \end{center}
  \caption{\label{fig:3} The same as Fig.~\ref{fig:2} but for the 10-20\%
    centrality bin.}
\end{figure}

\begin{figure}
  \begin{center}
  \epsfig{file=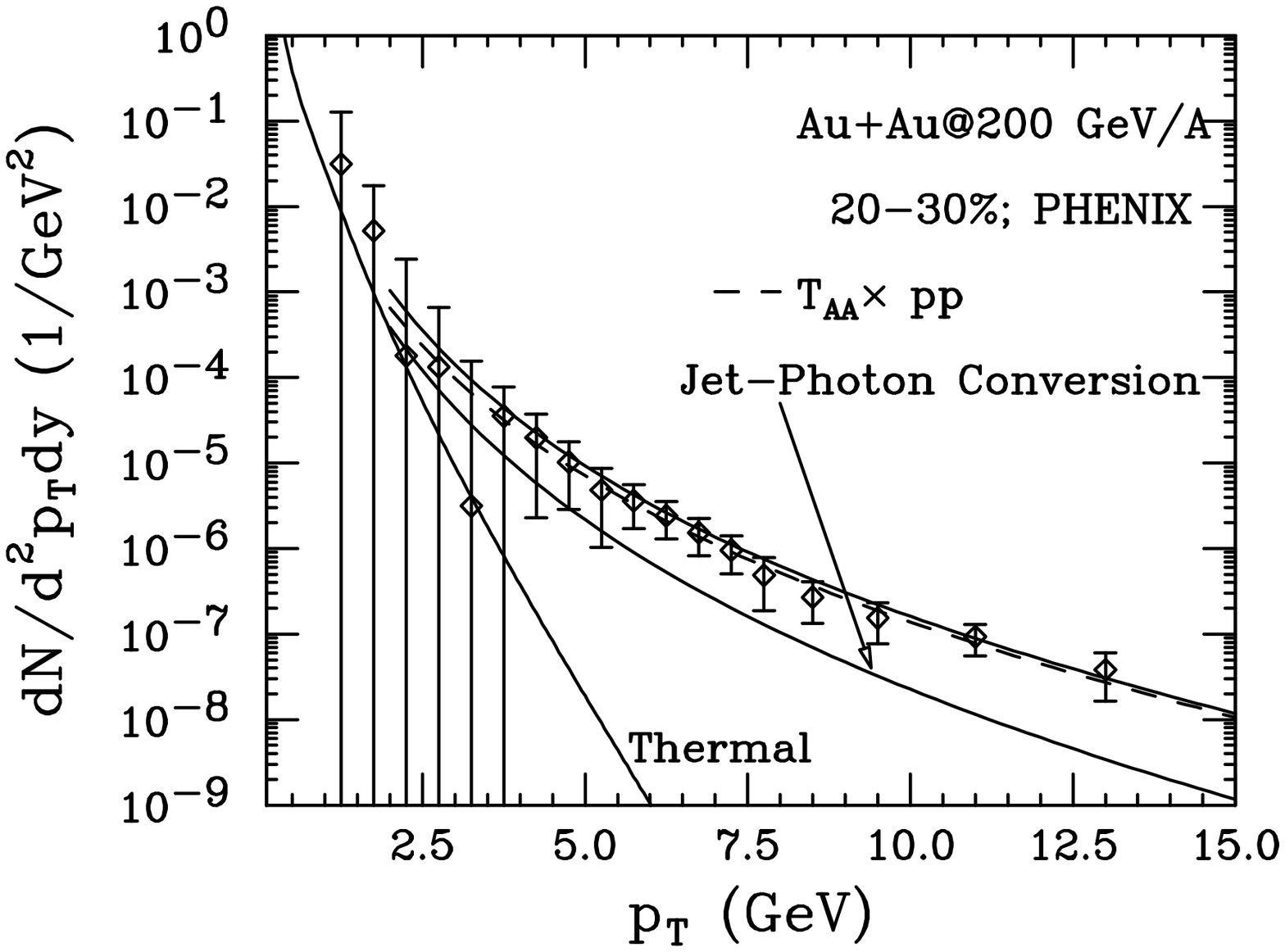,width=0.8\columnwidth}
  \end{center}
  \caption{\label{fig:4} The same as Fig.~\ref{fig:2} but for the 20-30\%
    centrality bin.}
\end{figure}

\begin{figure}
  \begin{center}
  \epsfig{file=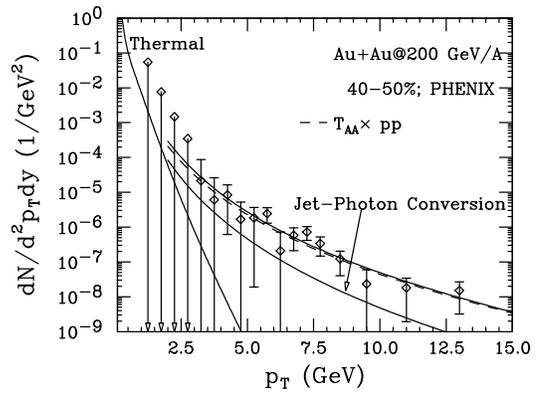,width=0.8\columnwidth}
  \end{center}
  \caption{\label{fig:5} The same as Fig.~\ref{fig:2} but for the 40-50\%
    centrality bin.}
\end{figure}

Figs.\ \ref{fig:2}, \ref{fig:3}, \ref{fig:4} and \ref{fig:5} present our 
results for photon yields at midrapidity in Au+Au collisions as a function 
of transverse momentum in four different centrality bins: 0-10\%, 10-20\%, 
20-30\% and 40-50\%. Both prompt photons and photons from jet-plasma 
interactions are shown, and the sum is compared to PHENIX data
\cite{phenix:05gammaauau}. For completeness we also show our result for 
thermal photons. The agreement with data is generally quite good for all 
centrality bins. This is also true for the other centrality bins, up to the
60-70\% bin,
which are not shown here.

As already pointed out in \cite{FMS:02}, the jet-photon conversion spectrum
is falling off with $p_T$ faster than the spectrum of prompt photons, a 
consequence of the additional hard momentum transfer with a propagator 
$\sim 1/p_T^2$. Interestingly the jet-photon conversion yield is as large as 
both the primary hard photon yield and the thermal yield around 
$p_T=2$ GeV/$c$ in central Au+Au collisions. Unfortunately, the quality of 
data in this region is still poor. Between 4 and 6 GeV/$c$, where the data 
provide a better constraint, jet-medium photons still add roughly a $50\%$ 
contribution on top of primary hard photons. 

It has been argued that photons from primary hard processes alone are 
sufficient to describe the Au+Au data \cite{phenix:05gammaauau,dEPe:05} and 
that no additional contribution is needed.
The first statement is correct, to a certain extent, as can 
be seen in Fig.~\ref{fig:2}. However, its value is limited by the size of the 
experimental error bars. We have demonstrated here that a calculation 
taking into account jet-photon conversion in a consistent manner also 
leads to a result which is compatible with present data. 
Caution should be excercised when isospin effects are omitted in the 
discussion.

Energy loss effects for jets before they convert into photons have been 
investigated by Turbide {\it et al.} \cite{TGJM:05}. The net effect on the
yield of photons from jet-medium interactions is found to be small, about 20\%.
This is because most photons are emitted at a time when the plasma is still 
hot. Jets have then traveled only a short distance through the plasma and
have not lost a significant amount of energy.

\begin{figure}
  \begin{center}
  \epsfig{file=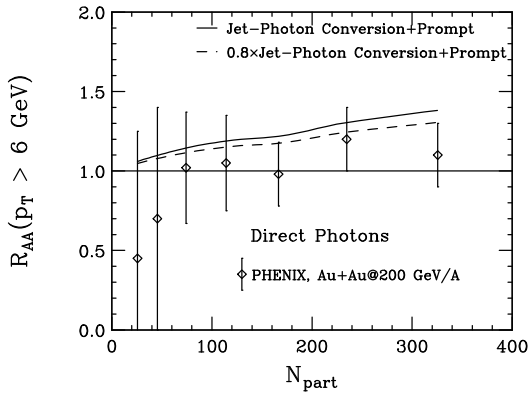,width=0.8\columnwidth}
  \end{center}
  \caption{\label{fig:6} The nuclear modification factor $R_{AA}$ for direct
    photons and
    $p_T >6$ GeV/$c$ as defined in (\ref{eq:3}) as a function of
    the number of participants $N_{\rm part}$ (solid line). We also show
    a calculation effectively taking into account energy loss of jets before
    convert into photons (dashed line). Data are taken from PHENIX 
    collaboration
    \cite{phenix:05gammaauau}.}
\end{figure}

The centrality dependence of the data is well described by our calculations. 
While direct photons scale with the number of collisions $N_{\rm coll}$,  
photons from jet-medium interactions exhibit a slightly stronger scaling. 
The relative importance of this process thus decreases in more peripheral 
collisions. This can also be seen in Fig.~\ref{fig:6} where we plot the 
nuclear modification factor 
\begin{equation}
  \label{eq:3}
  R_{AA} = \frac{dN_{AA}/dy|_{p_T > 6~{\rm GeV}/c}}{N_{\rm coll}\,
  dN_{pp}/dy|_{p_T > 6~{\rm GeV}/c}}
\end{equation}
calculated from the photon spectra integrated for $p_T > 6$ GeV/$c$ 
vs centrality. We show our calculation and a version taking into account 
an effective 20\% energy loss of jets before conversion into photons and 
compare with PHENIX data \cite{phenix:05gammaauau}. Due to the large error 
bars the data is compatible with 1 for central and midperipheral collisions,
in accordance with a binary collision scaling from $p+p$ collisions. Our 
calculation is compatible with the data as well, and predicts a slight rise
of $R_{AA}$ with centrality. Our results show that the importance of 
the contribution from jet-photon conversion to the total photon spectrum
grows below $p_T = 6$ GeV/$c$. A high statistics measurement of 
$R_{AA}$ in this region would be very useful.

In summary, we have calculated the photon spectrum in Au+Au collisions at 
RHIC energies resulting from primary hard photons, jets interacting with the 
medium and thermal radiation. The calculations are consistent with neutral 
pion and photon spectra measured in $p+p$. We obtain a good description of 
the $p_T$ dependence and centrality dependence of photon production in Au+Au.
The contribution from jet-photon conversions can be as large as 100\% (50\%)
that of primary hard photons at 2 (5) GeV/$c$ in accordance with data measured
by PHENIX. 

At the Large Hadron Collider (LHC), the importance of the jet-photon 
conversion process will be significantly enhanced. It has been shown that it
will be the dominant contribution below $p_T \approx 12$ GeV/$c$ 
\cite{FMS:02,TGJM:05}.

\begin{acknowledgments} This work was supported by DOE grants 
DE-FG02-87ER40328 and DE-FG02-05ER41367.
\end{acknowledgments}

\end{document}